# Vibronic coupling induced by fast Rabi oscillations for kinetic energy control in free atom

Andrei Ivanov* and Evgeniy Perlin

*Research Center "Information Optical Technologies",
ITMO University, Birzhevaya liniya, 14, Saint Petersburg, 199034, Russia*

Vibronic coupling effects usually manifest themselves in molecules and crystals rather than in unbound atoms. We theoretically demonstrate the existence of vibronic states in a moving two-level atom exposed to a strong electromagnetic wave. In this case, the Rabi oscillations of the electron density give rise to periodic displacements of the atomic center of mass with the Rabi frequency. The periodic displacements mix the Stark split electron levels and lead to the establishment of a channel for energy transfer in the photon-electron-center-of-mass system. Such a channel paves the way for fast control of the kinetic energy of a free atom with the use of the parameters of an electromagnetic wave. Thus, the system kinetic energy can be decreased or increased if the detuning between the laser pulse frequency and the optical transition is positive or negative, respectively. For actual values of the detuning, the pulse duration should be an order of magnitude longer than the lifetime of the excited atomic level in order to complete the energy transfer process.

## I. INTRODUCTION

Vibronic coupling (VC) effects exert a strong influence on the properties of a physical system in the case of the mixing of electronic and nuclear motions [1, 2]. Since these motions cannot be separated, the nonadiabaticity of electronic and nuclear coupling must be taken into account. This fully applies to the multi-atomic systems, which have rich vibrational spectra associated with the molecular dynamics. The optical properties of such systems are substantially determined by VC effects. For instance, the VC has been investigated in semiconductor quantum dots [3], molecular crystals [4], rare-earth ion-doped crystals [5, 6], nitrogen-vacancy centers in diamond [7], molecular dimers [8], bichromophoric molecules [9] and in other solid state and molecular systems. Unlike the systems listed above, free atoms do not have vibrational modes that play a key role in the VC effects. However, the situation can be changed by the influence of external fields.

Electric, magnetic, and electromagnetic (EM) fields are successfully used for cooling and trapping of neutral atoms and ions [10-13]. Field potentials of radio-frequency, magnetic, optical, and hybrid traps not only effectively confined atoms and ions but also force them to oscillate in the space with the characteristic frequency. The vibrational mode (VM) of the center-of-mass (c.m.) motion of the particle in the trap potential may be coupled via the laser to the electron states. This coupling of the external (vibrational) and internal (electronic) degrees of freedom allows realizing the laser cooling of a trapped particle to the motional ground state [14-16]. It worth stressing that such coupling of a trapped particle is described in the adiabatic approximation when electrons follow nucleus adiabatically. In this paper, we propose a mechanism for the formation of the VC induced by a strong EM field in an untrapped atomic-like system, taking into account the nonadiabaticity of electron and nucleus motions.

An important feature of light-matter interaction is the appearance of the electron density Rabi oscillations (ROs), which can be observed in experiments as the optical nutation [17]. In the case of the strong interaction, the optically coupled electron levels become split, which is the manifestation of the Autler-Townes effect, and the energy difference between Stark split sublevels is equal to the Rabi frequency in energy units. Changing the intensity of an EM field, the energy difference between split sublevels can be tuned to get the resonance with the vibrational motion of the c.m. of an atomic-like system. Such a scenario is used for getting the fast quantum gates for trapped cold ions [18], atomic [14] and ionic [19] cooling, nonclassical states of motion in a three-dimensional ion trap [20], measurement of the vibratory quantum state of a single ion [21], super-revivals in ion traps [22], and the vibronic Rabi resonances in harmonic and hard-wall ion traps [23]. These examples are related to the case when the Rabi frequency is less than the spontaneous decay rate of the excited electronic state. Recently, an analysis of the laser cooling of a trapped particle was performed for a much wider range of Rabi frequencies [24]. The analysis showed that an increase in the Rabi frequency gives rise to an increase in the effective cooling rate, but does not affect the final result of the cooling process. At the same time, in the case when the Rabi frequency, $\Omega$, is large compared to the spontaneous decay rate of the excited electronic state, $\gamma_0$, the electron density ROs need to be taken into account in the particle dynamics.

Let us consider an atom moving in space, which is suddenly exposed to a strong EM wave. The nonadiabatic switching on the interaction with the wave results in the fast ROs of electron density between the ground and excited atomic states if $\Omega \gg \gamma_0$. It is important to stress that the position of the atomic nucleus in space is changed if the atom is transferred to the excited electron state due to a change of the electron-nucleus interaction. In this case, the nucleus follows the electron nonadiabatically since it has a large mass compared to an electron mass. Thus, the ROs of the electron density lead to periodic displacements of the nucleus with the Rabi frequency that results in the VM emerging. Since the frequency of the VM is resonant to the frequency of the transition between Stark split electron sublevels, the VC is established. It is worth stressing that the resonance condition is always satisfied in this case.

The physical mechanism of the VC formation is based on the nonadiabaticity of electron and nucleus motions. Thus, two types of nonadiabaticity are included in the consideration for the EM field-atom and the electron-nucleus interactions. It should be emphasized that the considered physical situation can be interpreted as a manifestation of the pseudo-Jahn-Teller effect [25] in the electron-photon system. The analogous vibronic mechanism for two-level systems linked to the phonon reservoir was recently proposed for the case of internal degrees of freedom only [26]. The proposed in the current paper mechanism of the VC formation plays a key role in energy transfer in the photon-electron-vibrational system since it is determined the coupling between atomic internal and external degrees of freedom. With the use of the parameters of an EM field, fast control of the kinetic energy of a free atom can be implemented. It can be practically useful for the steering of atomic beams, acceleration or deceleration of atoms, and creation of atomic mirrors. The paper is organized as follows. In Sec. 2, we present the appearance of a VM of the nucleus induced by an EM wave and photoinduced vibronic states formed by the interaction of electrons with the VM. In Sec. 3, we calculate the decay rate for the dressed state (DS) basis in the presence of the electron-vibrational interaction and describe the physical mechanism of energy transfer in the electron-photon-vibrational system.

## II. PHOTOINDUCED VIBRONIC COUPLING

### A. Vibrational mode of two-level atomic-like system

The simplest model, which theoretically reveals the ROs, is a two-level system (TLS) that interacts with the monochromatic EM wave. Despite its simplicity, this model still attracts researchers by a wide range of physical conditions to be taking into account. In our consideration, we model an atom moving in space and interacting with a strong monochromatic EM wave as a TLS, taking into account the mixing of electron and nucleus motions. We are interested in the field-matter interaction condition when $\Omega \gg \gamma_0$. As it is mentioned before, the fast ROs give rise to changing of the position of the electron relative to the atomic nucleus with the Rabi frequency. A change of the electron position leads to a displacement of the nuclear position. Such displacement can be linked to a change in the potential energy of the atomic system. To include in our consideration the change of the potential energy, we present the operator of electron-nucleus interaction as an expansion of the Taylor's series in the small displacements of the nuclear position in the following form

$$\hat{W}(\mathbf{r}_e, \mathbf{r}_n) = \hat{W}(\mathbf{r}_e, \mathbf{r}_n^0) + (\mathbf{r}_n - \mathbf{r}_n^0) \frac{\partial \hat{W}(\mathbf{r}_e, \mathbf{r}_n^0)}{\partial \mathbf{r}_n}$$
$$+ \frac{(\mathbf{r}_n - \mathbf{r}_n^0)^2}{2} \frac{\partial^2 \hat{W}(\mathbf{r}_e, \mathbf{r}_n^0)}{\partial \mathbf{r}_n^2} + .. \quad (1a)$$

Here $\mathbf{r}_e$ and $\mathbf{r}_n$ are the position operators of the atomic electron and nucleus, respectively; $\mathbf{r}_n^0$ is the nuclear position associated with the ground electron state of the atom. Then, we take into account the fact that the nuclear mass is much larger than the electron mass, and the positions of the c.m. and the nucleus are close to each other. For instance, in the case of the simple model of the Hydrogen atom, such relationship is $\mathbf{r}_n = \mathbf{R} - \mathbf{r} m_e / M \approx \mathbf{R}$, where $M$ and $m_e$ are the full atomic mass and the electron mass, respectively; $\mathbf{R}$ and $\mathbf{r} = \mathbf{r}_e - \mathbf{r}_n^0$ are the position operators of the c.m. and an electron with respect to the nucleus, respectively. This assumption is close to the description of c.m. dynamics in Ref. 27. Using this conclusion, we rewrite Eq. (1a) as a function that depends on the position of the c.m. and the position of an electron with respect to the nucleus

$$\hat{W}(\mathbf{r}, \mathbf{R}) = \hat{W}(\mathbf{r}) + \Delta \mathbf{R} \hat{W}^{(1)}(\mathbf{r}) + \frac{\Delta \mathbf{R}^2}{2} \hat{W}^{(2)}(\mathbf{r}) + ...,$$
$$\hat{W}^{(1)}(\mathbf{r}) = \frac{\partial \hat{W}(\mathbf{r}_e - \mathbf{r}_n^0)}{\partial \mathbf{r}_n}, \ \hat{W}^{(2)}(\mathbf{r}) = \frac{\partial^2 \hat{W}(\mathbf{r}_e - \mathbf{r}_n^0)}{\partial \mathbf{r}_n^2},$$
$$\Delta \mathbf{R} = \mathbf{R} - \mathbf{R}_0, \quad (1b)$$

where $\Delta \mathbf{R}$ is the displacement operator and $\mathbf{R}_0$ corresponds with $\mathbf{r}_n^0$. The considered situation is depicted in Fig. 1 where the nuclear position is fixed.

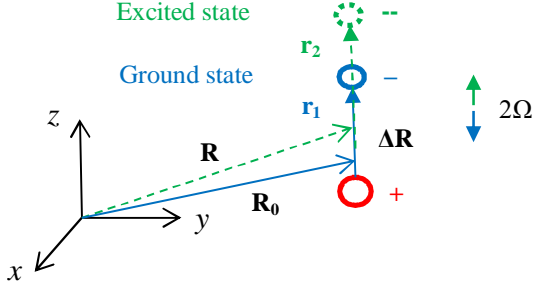

FIG. 1. (Color online) Oscillations of the atomic center of mass with the Rabi frequency in the laboratory frame in the case of the fixed nuclear position.

Thus, the Hamiltonian of the system can be written as

$$\hat{H} = \hat{H}_1(\mathbf{r}) + \hat{H}_2(\mathbf{r}, \mathbf{R}),$$
$$\hat{H}_1 = \frac{\varepsilon}{2}\hat{\sigma}_z + \hbar\omega_0\left(\hat{c}^+\hat{c} + \frac{1}{2}\right) - \hbar g_0\left(\hat{\sigma}_+\hat{c} + \hat{c}^+\hat{\sigma}_-\right),$$
$$\hat{H}_2 = \hat{T}_K(\mathbf{R}) + \Delta\hat{W}(\mathbf{r},\mathbf{R}),$$
$$\Delta\hat{W}(\mathbf{r},\mathbf{R}) = \hat{W}(\mathbf{r},\mathbf{R}) - \hat{W}(\mathbf{r}). \quad (2)$$

Here, $\varepsilon$ is the energy difference between the ground and excited electron levels; $\sigma_z$, $\sigma_+$, and $\sigma_-$ are the Pauli operators associated to the TLS with the ground $|\varphi_2\rangle$ and the excited $|\varphi_1\rangle$ electron states which are called bare states (BSs); $c$ and $c^+$ are the annihilation and creation operators of the EM wave of frequency $\omega_0$; $g_0$ is the coupling strength of the electron-photon interaction; $\hat{T}_K$ is the operator of the c.m. kinetic energy. In Eq. (2), the total Hamiltonian is divided into two parts. The first part, $\hat{H}_1$, contains the electron and photon subsystems and their interaction in the dipole and the rotating wave approximations. That is the Jaynes-Cummings Hamiltonian (JCH), which we write in the ordinary form. It should be noted that the first term of the series of Eq. (1b) is included in the JCH. The second part, $\hat{H}_2$, contains the energy contributions associated with the c.m. motion.

In Eq. (2), we omit the EM wave dependence on the c.m. position operator since we assume that the EM wave propagates in the orthogonal direction to the motion of the atom. Such geometry of the problem is chosen to concentrate on the features of the wave action of light on an atomic system and exclude corpuscular nature of light associated with the wave vector of the EM wave. In general case, chosen geometry is not mandatory, and mutual configuration of the EM wave and the atomic system can be arbitrary. However, taking into account the EM wave dependence on the c.m. position complicates calculations and can make the results unclear.

Additionally, we assume that a change in the atomic momentum due to the *corpuscular light pressure* [10, 11] is negligible, while the *wave light pressure* considered in the current paper changes the atomic momentum substantially. For that reason, we regard the case in which the dynamical Stark shift is much larger than the Doppler shift. Thus, the latter shift is discarded from the consideration. In this case, the Rabi oscillation frequency has to be large compared to the spontaneous decay rate of an excited electronic level. For average decay rate ($10^8 - 10^9$ Hz) of excited levels of atomic systems (allowed transitions), it means that the value of the Rabi frequency has to be higher than 10 GHz (intensity of the external EM wave is higher than 10 kW/cm$^2$). For optical radiation ($10^{15}/2\pi$ Hz) and average velocity of an atomic beam of $5\times10^4$ cm/s, the Doppler shift is about 1 GHz.

Then, we define the approximate eigenfunction of the Hamiltonian (2) with the operator (1b) in the adiabatic (for the electron-nucleus interaction) basis

$$|\Psi(\mathbf{r},\mathbf{R})\rangle = |\psi(\mathbf{r},\mathbf{R})\rangle|\Phi(\mathbf{R})\rangle. \quad (3)$$

The electron wave function $|\psi(\mathbf{r},\mathbf{R}_0)\rangle$ is the solution of the JCH for the c.m. position $\mathbf{R}_0$. This solution is well known and can be written in three different bases, namely, in the basis of bare and dressed electron states [28, 29] and in the basis of eigenstates of the parity operator [26, 30]. For our consideration, we choose the DS basis and write the solution for $\hat{H}_1$ with the initial condition in the case when only the ground electron state $|\varphi_2\rangle$ is populated in the following form

$$|\psi(\mathbf{r},\mathbf{R}_0,t),m\rangle$$
$$= c_m u_- |\xi_1(\mathbf{r},\mathbf{R}_0,t),m\rangle + c_m u_+ |\xi_2(\mathbf{r},\mathbf{R}_0,t),m\rangle,$$
$$u_\pm = \sqrt{\frac{1}{2}\left(1 \pm \frac{\delta}{2\Omega}\right)}, \quad 2\Omega = \sqrt{\delta^2 + 4|g_0|^2(m+1)},$$
$$\delta = \hbar^{-1}\varepsilon - \omega_0. \quad (4)$$

Here, $m$ is the number of photons; $c_m$ is the initial probability amplitude for the state with $m$ photons; $\Omega$ is the off-resonance Rabi frequency; $\delta$ is the detuning of the EM wave frequency with respect to the electron transition frequency. It should be stressed that the wave function in Eq. (4) is defined for the instant start of the electron-photon interaction in the nonadiabatic (for this type of interaction) case [28, 29]. For the convenience of further consideration, the DSs can be presented in the BS basis

$$\left|\xi_1(\mathbf{r},\mathbf{R}_0,t),m\right\rangle = u_-e^{-i\left(\Omega+\frac{\delta}{2}\right)t}\left|\varphi_2(\mathbf{r},\mathbf{R}_0,t),m+1\right\rangle$$
$$-u_+e^{-i\left(\Omega-\frac{\delta}{2}\right)t}\left|\varphi_1(\mathbf{r},\mathbf{R}_0,t),m\right\rangle,$$
$$\left|\xi_2(\mathbf{r},\mathbf{R}_0,t),m\right\rangle = u_+e^{i\left(\Omega-\frac{\delta}{2}\right)t}\left|\varphi_2(\mathbf{r},\mathbf{R}_0,t),m+1\right\rangle$$
$$+u_-e^{i\left(\Omega+\frac{\delta}{2}\right)t}\left|\varphi_1(\mathbf{r},\mathbf{R}_0,t),m\right\rangle. \quad (5)$$

Then, we find the solution for the Schrödinger equation

$$\left(\hat{H}_1(\mathbf{r})+\Delta\hat{W}(\mathbf{r},\mathbf{R})\right)\left|\psi(\mathbf{r},\mathbf{R})\right\rangle = E(\mathbf{R})\left|\psi(\mathbf{r},\mathbf{R})\right\rangle \quad (6)$$

as a linear combination of the DSs

$$\left|\psi,m\right\rangle = d_1(t)\left|\xi_1,m\right\rangle + d_2(t)\left|\xi_2,m\right\rangle \quad (7)$$

with the initial conditions for the positive detuning δ in correspondence with Eq. (4)

$$d_1(0)=c_m u_-, \ d_2(0)=c_m u_+. \quad (8)$$

Using the unitary transformation $U(t)=\exp\left[-i\hbar^{-1}H_1 t\right]$ to get the interaction representation, the following differential system for the coefficients can be readily obtained

$$i\hbar\frac{\partial}{\partial t}D_1(t) = \left(\Delta\mathbf{R}\,F_{12}^{(1)}+\frac{\Delta\mathbf{R}^2}{2}F_{12}^{(2)}\right)e^{i2\Omega't}D_2(t),$$
$$i\hbar\frac{\partial}{\partial t}D_2(t) = \left(\Delta\mathbf{R}\,F_{21}^{(1)}+\frac{\Delta\mathbf{R}^2}{2}F_{21}^{(2)}\right)e^{-i2\Omega't}D_1(t),$$
$$D_j(t) = \exp\left\{\frac{it}{\hbar}\left(\Delta\mathbf{R}\,F_{jj}^{(1)}+\frac{\Delta\mathbf{R}^2}{2}F_{jj}^{(2)}\right)\right\}d_j(t),$$
$$\Omega' = \Omega - \frac{\Delta\mathbf{R}}{2\hbar}\left(F_{22}^{(1)}-F_{11}^{(1)}\right)-\frac{\Delta\mathbf{R}^2}{4\hbar}\left(F_{22}^{(2)}-F_{11}^{(2)}\right),$$
$$F_{11}^{(j)}=u_+^2 W_{11}^{(j)}+u_-^2 W_{22}^{(j)},\ F_{22}^{(j)}=u_-^2 W_{11}^{(j)}+u_+^2 W_{22}^{(j)},$$
$$F_{12}^{(j)}=u_+u_-\left(W_{22}^{(j)}-W_{11}^{(j)}\right),$$
$$W_{ij}^{(1)}=\left\langle\varphi_i(\mathbf{r},\mathbf{R}_0)\left|\frac{\partial\hat{W}(\mathbf{r},\mathbf{R}_0)}{\partial\mathbf{R}}\right|\varphi_j(\mathbf{r},\mathbf{R}_0)\right\rangle,$$
$$W_{ij}^{(2)}=\left\langle\varphi_i(\mathbf{r},\mathbf{R}_0)\left|\frac{\partial^2\hat{W}(\mathbf{r},\mathbf{R}_0)}{\partial\mathbf{R}^2}\right|\varphi_j(\mathbf{r},\mathbf{R}_0)\right\rangle,\ i,j=1,2. \quad (9)$$

In the solution obtained, we use the so-called harmonic approximation for the operator $\Delta\hat{W}$, saving the terms up to the second order with respect to the small displacements of the atomic c.m. position. The solution of the system gives the wave function which can be written, also like DSs, in different bases: (i) in the BS basis, (ii) in the DS basis, (iii) in the basis of the doubly dressed states (DDSs), and in the basis of the parity operator states (POSs). In the third case, the electron states are sequentially dressed by the electron-photon and electron-nucleus interactions. The electron-photon interaction mixes the upper and lower BSs and results in the conventional AC Stark effect with splitting of the BSs. The electron-nucleus interaction mixes the DSs and results in splitting of the DSs (Fig. 2). Thus, the first interaction is responsible for the Stark broadening while the latter one, as it follows from our consideration, leads to emerging of the vibrational broadening of spectral lines. In the DDS basis, the obtained wave function can be presented as

$$\left|\psi(\mathbf{r},\mathbf{R},t),m\right\rangle = c_m V_1 \left|\zeta_1(\mathbf{r},\mathbf{R},t),m\right\rangle + c_m V_2 \left|\zeta_2(\mathbf{r},\mathbf{R},t),m\right\rangle,$$
$$\left|\zeta_1(\mathbf{r},\mathbf{R},t),m\right\rangle = \left(v_+e^{i\Omega t}\left|\xi_1(\mathbf{r},\mathbf{R}_0,t),m\right\rangle\right.$$
$$\left.+v_-e^{-i\Omega t}\left|\xi_2(\mathbf{r},\mathbf{R}_0,t),m\right\rangle\right)e^{-i(\Omega_1+\Omega'')t},$$
$$\left|\zeta_2(\mathbf{r},\mathbf{R},t),m\right\rangle = \left(-v_-e^{i\Omega t}\left|\xi_1(\mathbf{r},\mathbf{R}_0,t),m\right\rangle\right.$$
$$\left.+v_+e^{-i\Omega t}\left|\xi_2(\mathbf{r},\mathbf{R}_0,t),m\right\rangle\right)e^{i(\Omega_1-\Omega'')t},$$
$$V_1=\frac{u_-v_++u_+v_-}{\sqrt{2}},\ V_2=\frac{u_+v_+-u_-v_-}{\sqrt{2}},$$
$$\Omega''\equiv\hbar^{-1}\left(\Delta\mathbf{R}\left(F_{11}^{(1)}+F_{22}^{(1)}\right)+\frac{1}{2}\Delta\mathbf{R}^2\left(F_{11}^{(2)}+F_{22}^{(2)}\right)\right),$$
$$v_\pm=\sqrt{\frac{1}{2}\pm\frac{\Omega'}{2\Omega_1}},\ \Omega_1=\sqrt{\Omega'^2+\Omega_0^2},\ \Omega_0=\Delta\mathbf{R}\frac{F_{12}^{(1)}}{\hbar}. \quad (10)$$

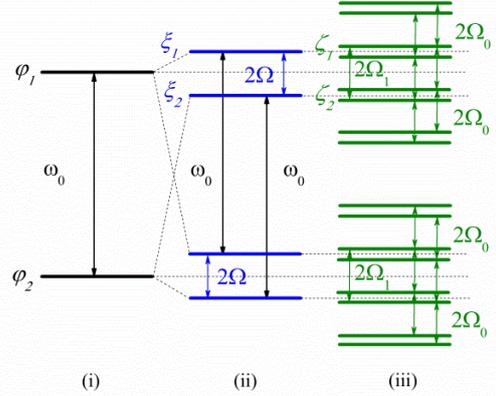

FIG. 2. (Color online) Formation of the electron level structure in the two-level system: (i) BSs with an electromagnetic field of frequency $\omega_0$; (ii) DSs split by the electron-photon interaction; (iii) DDSs and POSs formed by the electron-nucleus interaction. The differences in energy between the levels are defined in the text.

In the calculations, we assume that the vectors $\Delta\mathbf{R}$ and $\mathbf{r}$ are parallel (see Fig. 1) that means parallelism of a dipole moment of the atom and the displacement vector of the atomic c.m. Eqs. (10) show that the electron population in the TLS oscillates between the DSs with the frequency $\Omega_1$.

Therefore, the adiabatic potential of Eq. (6) takes the form

$$\hat{E}(\mathbf{R}) = \hbar\Omega_1(\mathbf{R})\hat{\sigma}_z + \hbar\Omega''(\mathbf{R})\hat{\sigma}_0. \tag{11}$$

In order to define the wave function of Eq. (3), we have to solve the Schrödinger equation for Hamiltonian, which describes the c.m. motion

$$\left(\hat{T}_K(\mathbf{R}) + \hat{E}(\mathbf{R})\right)|\Phi(\mathbf{R})\rangle = \Lambda|\Phi(\mathbf{R})\rangle. \tag{12}$$

Hamiltonian (12) is derived for the case in which the so-called nonadiabatic operator is neglected [1]. In the next step of our consideration, we represent Eq. (11) using a series expansion of the square root function in the $\Omega_1$ up to the second order with respect to the small displacements of the c.m. position

$$\hat{E}(\mathbf{R}) = \left(\hbar\Omega + \frac{1}{\hbar\Omega}\Delta\mathbf{R}^2\left|F_{12}^{(1)}\right|^2\right)\hat{\sigma}_z$$
$$+ \left(\Delta\mathbf{R} F^{(1)} + \frac{1}{2}\Delta\mathbf{R}^2 F^{(2)}\right)\hat{\sigma}_0, \tag{13}$$

where $F^{(1)}$ and $F^{(2)}$ are the matrix elements defined in Eq. (9). Then, we assume that an atom moves along the $z$ coordinate axis in the laboratory frame, and an EM wave, which is linearly polarized in the $z$-direction, propagates along the $y$-axis. Taking into account these assumptions and utilizing the displacement transformation [31]

$$\hat{S} = \exp\left\{\lambda\frac{\partial}{\partial z}\right\}, \tag{14}$$

we transform Eq. (12) to the equation for the quantum harmonic oscillator

$$\left(\frac{\partial^2}{\partial z^2} - \frac{M\tilde{\Omega}_{jj}^2}{\hbar^2}z^2 + \frac{2M}{\hbar^2}\tilde{\Lambda}_j\right)|\tilde{\Phi}_j(z)\rangle = 0,$$

$$|\tilde{\Phi}(z)\rangle = \hat{S}^+|\Phi(z)\rangle, \quad M\tilde{\Omega}_{jj}^2 = \frac{\left|F_{12}^{(1)}\right|^2}{\hbar\Omega} - (-1)^j F_{jj}^{(2)},$$

$$\tilde{\Lambda}_j = \Lambda_j + \frac{\lambda_j F_{jj}^{(1)}}{2} - (-1)^j \hbar\Omega, \quad \lambda_j = \frac{F_{jj}^{(1)}}{M\tilde{\Omega}_{jj}^2}. \tag{15}$$

Here, $M$ is the mass of the atom. It should be stressed that Eqs. (15) describe the harmonic oscillations of the c.m. position at two electron levels with different frequencies $\tilde{\Omega}_{jj}$. Such a situation gives rise to the nonorthogonality of two sets of the wave functions $|\tilde{\Phi}_j(z)\rangle$. However, the orthogonality of the functions of Eq. (3), associated with different states, is provided by the orthogonality of the electron functions. Additionally, the transformation of Eq. (13) with the use of Eq. (14) leads to displacements of the original wave functions $|\Phi_j(z)\rangle$ by the different amounts of $\lambda_j$ for two electron levels. In further consideration, we neglect the effects related to the frequency difference and assume that $\tilde{\Omega}_{11} = \tilde{\Omega}_{22} \approx 2\Omega_0$. This assumption follows from Eqs. (15) and from the last equation of Eqs. (10) if we define the amplitude of the displacement of the atomic c.m. position as $\sqrt{\hbar/4\Omega M}$. Therefore, the photoinduced vibronic state takes the form

$$|\Psi(t)\rangle = \sum_m\sum_n\frac{c_m}{\sqrt{2}}\left(|\zeta_1(t),m,n\rangle + |\zeta_2(t),m,n\rangle\right),$$

$$E_j = \hbar\omega_0(m+1) - (-1)^j\hbar\left\{2\Omega_0\left(n+\frac{1}{2}\right) + \Omega\right\} - \frac{\lambda_j}{2}. \tag{16}$$

In Eqs. (16), we use the second-quantization representation for the photoinduced VM with the number of quanta $n$. Thus, the emerging VM, based on the electron-nuclear interaction, leads to the splitting of the electron levels once again (Fig. 2). The existence of the linear term in the adiabatic potential of Eq. (13) is responsible for not only displacements of the wave functions $|\Phi_j(z)\rangle$ but also for the change in the eigenvalues of Eq. (12). The last term in Eq. (16) for the DDS energies plays the same role as the recoil energy in the case of the corpuscular light pressure description [10, 11].

### B. Vibronic states in electron-photon subsystem

In Subsec. 1, we take into account the electron-nucleus interaction in order to determine the wave functions, which depend on the c.m. position. Here, we consider this interaction to obtain the electron wave functions. To this end, we have to define the electron Hamiltonian of the considered problem.

As can be seen from Subsec. 1, the fast ROs of the electron density under the action of the EM wave give rise to the appearance of the c.m. VM and the coupling of this VM with the electron states. Utilizing these results, we can rewrite the system Hamiltonian [see Eq. (2)] in the second quantization formalism as follows

$$\hat{H} = \hat{H}_1 + 2\hbar\Omega_0\left(\hat{a}^+\hat{a} + \frac{1}{2}\right) + \hbar\Omega_0(\hat{a}^+\hat{\sigma}'_- + \hat{\sigma}'_+\hat{a}),$$

$$\hat{H}_1 = \hbar\Omega\hat{\sigma}'_z + \hbar\omega_0\left(\hat{c}^+\hat{c} + 1\right). \tag{17}$$

Here, $\hat{a}$ and $\hat{a}^+$ are the annihilation and creation operators of the VM; $\hat{\sigma}'_i$ are the electron (Pauli) operators in the DS basis. Strictly speaking, these electron operators differ from the operators in Eq. (1) and are related to them by a canonical transformation (see, for instance, [32]). Hamiltonian (17) is the JCH describing the interaction of the c.m.VM with the electron subsystem dressed by an EM wave. Since this Hamiltonian commutes with the parity operator, the system wave function can be presented as a linear combination of the eigenstates of this operator [30]

$$|\psi(t),m,n\rangle = D_n^+(t)|\eta_+,m,n\rangle + D_n^-(t)|\eta_-,m,n\rangle,$$
$$\hat{\Pi}|\eta_+,m,n\rangle = |\eta_+,m,n\rangle, \quad \hat{\Pi}|\eta_-,m,n\rangle = -|\eta_-,m,n\rangle. \quad (18)$$

Here, $\hat{\Pi} = -\hat{\sigma}'_z(-1)^{\hat{a}^+\hat{a}}$ is the parity operator; $|\eta_-,m,n\rangle$ and $|\eta_+,m,n\rangle$ are the odd and even parity states, respectively. In turn, the solution for the Schrödinger equation with the Hamiltonian (17) is seeking in the DS basis in the following form [26]

$$|\eta_-,m,n\rangle = D_{2n}^-(t)|\xi_1,m,2n\rangle + D_{2n+1}^-(t)|\xi_2,m,2n+1\rangle,$$
$$|\eta_+,m,n\rangle = D_{2n-1}^+(t)|\xi_1,m,2n-1\rangle + D_{2n}^+(t)|\xi_2,m,2n\rangle. \quad (19)$$

The expansion coefficients in Eq. (19), $D_n^\pm$, are determined with the initial conditions of Eq. (8) for even and odd values of quanta $n$ in the VM at initial time moment separately. Finally, the solution can be written in the following form

$$|\psi_+,m,n\rangle = c_m d_{2n}\left(u_-|\eta_-,m,2n+1\rangle + u_+|\eta_+,m,2n\rangle\right),$$
$$|\psi_-,m,n\rangle = c_m d_{2n+1}\left(u_+|\eta_-,m,2n+1\rangle + u_-|\eta_+,m,2n\rangle\right),$$
$$|\eta_-,m,2n+1\rangle = \sqrt{2}\cos\left(\sqrt{2n+1}\Omega_0 t\right)|\xi_1,m,2n\rangle$$
$$-\sqrt{2}i\sin\left(\sqrt{2n+1}\Omega_0 t\right)|\xi_2,m,2n+1\rangle$$
$$= \frac{1}{\sqrt{2}}\left(|\zeta_1,m,2n+1\rangle + |\zeta_2,m,2n+1\rangle\right),$$
$$|\eta_+,m,2n\rangle = \sqrt{2}\cos\left(\sqrt{2n}\Omega_0 t\right)|\xi_2,m,2n\rangle$$
$$-\sqrt{2}i\sin\left(\sqrt{2n}\Omega_0 t\right)|\xi_1,m,2n-1\rangle$$
$$= \frac{1}{\sqrt{2}}\left(|\zeta_1,m,2n\rangle - |\zeta_2,m,2n\rangle\right),$$
$$|\zeta_1,m,2n\rangle = \frac{1}{\sqrt{2}}\left(|\xi_1,m,2n-1\rangle + |\xi_2,m,2n\rangle\right)e^{-\frac{i}{\hbar}E_1 t},$$
$$|\zeta_2,m,2n\rangle = \frac{1}{\sqrt{2}}\left(|\xi_1,m,2n-1\rangle - |\xi_2,m,2n\rangle\right)e^{-\frac{i}{\hbar}E_2 t},$$
$$\hbar^{-1}E_{1,2} = \omega_0(m+1) \pm 2\Omega_0\left(n+\frac{1}{2}\right) \pm \sqrt{2n}\Omega_0. \quad (20)$$

Here, $|\psi_+,m,n\rangle$ and $|\psi_-,m,n\rangle$ are the system states for even and odd values of vibrational quanta at initial time moment; $d_n$ is the initial probability amplitude for the vibrational state with $n$ quanta. The solution is presented in the POS, DDS, and DS bases (see Fig 2). Additionally, Eqs. (20) are obtained, taking into account the resonance condition ($\Omega \approx \Omega_0$, $v_+ = v_- = 2^{-1/2}$). Eqs. (20) show that the solution for the POSs reveals the situation when the electron density oscillates between the DSs with the frequency $\Omega_1 \approx \sqrt{2n}\Omega_0$. Moreover, the DSs have initially different populations [see Eq.(8)], which become equal due to the electron-vibrational interaction.

The obtained results describe the mechanism of the photoinduced VC formation in the atomic-like TLS as follows. The fast ROs of the electron density between the BSs periodically displace the atomic c.m. position with the Rabi frequency and give rise to the appearance of a VM in the translational movement of the atomic system. The vibrational amplitude depends on the displacement module, electron-nucleus and electron-photon coupling constants [see Eqs. (9)]. Since the frequency of the vibrations is equal to the frequency of the electron transition between the DSs, the forced VM mixes the electron DSs. That is the reason for the electron density oscillations between these states and defines the back action of the vibrations on the electron density. Thus, the interaction of Eq. (1b) results in the appearance of the VC in the electron-photon subsystem.

### III. CONTROL OF KINETIC ENERGY BY LASER PULSES

#### A. Decay rate for dressed state basis in presence of electron-vibrational interaction

The establishment of the VC in the atomic-like TLS paves the way for fast control of the kinetic energy of an atom with the use of the parameters of an EM wave. For taking into account the energy transfer in the system, we have to utilize the density matrix formalism and regard the link of the system with a reservoir. It is important to emphasize that, in the case under consideration, there must be two reservoirs that are associated with the continuum of vacuum EM modes and the momentum space. However, coupling between TLS and these reservoirs is provided by only one decay process, namely the process of spontaneous emission. It becomes obvious if we take into account the atomic c.m. momentum change due to the photon emission (recoil effect [33]). Spontaneously emitting photons randomly change not only the modulus but also the direction of the full atomic c.m. momentum.

Now, we consider the interaction of the system with the vacuum EM modes in the presence of the electron-vibrational interaction in the system. To do this, we write the operator of the system-reservoir interaction in the interaction picture

$$\hat{V}_{SR} = \sum_{\mathbf{k}} \hbar g_{\mathbf{k}} \exp\left\{\frac{it}{\hbar}(\hat{H}+\hat{H}_R)\right\} \hat{B}_{\mathbf{k}} \exp\left\{-\frac{it}{\hbar}(\hat{H}+\hat{H}_R)\right\},$$
$$\hat{H}_R = \hbar\omega_{\mathbf{k}}\left(\hat{b}_{\mathbf{k}}^+\hat{b}_{\mathbf{k}} + \frac{1}{2}\right), \quad \hat{B}_{\mathbf{k}} = \hat{b}_{\mathbf{k}}^+\hat{\sigma}_- + \hat{\sigma}_+\hat{b}_{\mathbf{k}}. \quad (21)$$

Here, $\hat{b}_{\mathbf{k}}$ and $\hat{b}_{\mathbf{k}}^+$ are the annihilation and creation operators for each vacuum EM mode with the wave vector $\mathbf{k}$. It must be stressed that the electron operators for the operator $\hat{B}_{\mathbf{k}}$ in Eq. (21) are written in the BS basis. Since the system

Hamiltonian includes the electron-vibrational interaction, the exponential operators in Eq. (21) contain noncommuting terms that can be disentangled by the use of the Feynman formula

$$\exp\{-i\hbar^{-1}(\hat{H}_0 + \Delta\hat{W})t\}$$
$$= \exp\{-i\hbar^{-1}\hat{H}_0 t\}\hat{T}\exp\{-i\hbar^{-1}\int dt'\Delta\hat{W}(t')\},$$
$$\Delta\hat{W}(t) = \exp\{-i\hbar^{-1}\hat{H}_0 t\}\Delta\hat{W}\exp\{i\hbar^{-1}\hat{H}_0 t\}. \quad (22)$$

In the considered case, the operators in Eq. (22) have the following form

$$\hat{H}_0 = \hat{H}_R + \hat{H}_1 + 2\hbar\Omega_0\left(\hat{a}^+\hat{a} + \frac{1}{2}\right),$$
$$\Delta\hat{W}(t) = \hbar\Omega_0\left(\hat{a}^+\hat{\sigma}'_- e^{-2i(\Omega-\Omega_0)t} + \hat{\sigma}'_+\hat{a} e^{2i(\Omega-\Omega_0)t}\right). \quad (23)$$

For the reason that we assume the case of an exact resonance between the electron transition in the DS basis and the frequency of the VM, we can replace time-ordered operators with exponential ones. After some calculations, Eq. (21) can be rewritten as follows

$$\hat{V}_{SR} = \sum_{\mathbf{k}} \hbar g_{\mathbf{k}} \hat{D}(i\Omega_0 t)\hat{B}_{\mathbf{k}}(t)\hat{D}(-i\Omega_0 t),$$
$$\hat{B}_{\mathbf{k}}(t) = \hat{b}^+_{\mathbf{k}}\hat{\sigma}_- \exp\{-i\delta_{\mathbf{k}}t\} + \hat{\sigma}_+ \hat{b}_{\mathbf{k}} \exp\{i\delta_{\mathbf{k}}t\},$$
$$\delta_{\mathbf{k}} = \hbar^{-1}\varepsilon - \omega_{\mathbf{k}}, \quad \hat{D}(i\Omega_0 t) = \exp\{i\Omega_0 t(\hat{a}^+\hat{\sigma}'_- + \hat{\sigma}'_+\hat{a})\}. \quad (24)$$

Here, $\hat{D}(\alpha) \equiv \alpha\hat{a}^+ - \alpha^*\hat{a}$ is the displacement operator. It should be stressed that the DDSs are the eigenfunctions of the displacement operator in the considered system

$$\hat{D}(\pm i\Omega_0 t)|\zeta_i, m, n\rangle = e^{\pm i\Omega_0 t\sqrt{n+1}}|\zeta_i, m, n\rangle. \quad (25)$$

Then, we use Eq. (24) and the master equation in the Lindblad form [34] for the reduced density matrix of the system in order to describe the spontaneous emission in the TLS, following the procedure from Ref. 28.

Thus, we take the master equation in the form

$$\frac{\partial \hat{\rho}_S(t)}{\partial t} \approx \sum_m \text{Tr}_R \left\{ -\frac{i}{\hbar}[\hat{V}_{SR}(t), \hat{\rho}_S^{(m)}(t) \otimes \hat{\rho}_R(0)] \right.$$
$$\left. -\frac{1}{\hbar^2}\int_0^t dt'[\hat{V}_{SR}(t),[\hat{V}_{SR}(t'), \hat{\rho}_S^{(m)}(t) \otimes \hat{\rho}_R(0)]] \right\}. \quad (26)$$

In Eq. (26), $\hat{\rho}_S$ and $\hat{\rho}_R$ are the density matrix operators for the system and reservoir, respectively; $\text{Tr}_R\{\}$ means the process of taking a trace over reservoir states; $\otimes$ stands for a direct product of operators; and the summation is performed over EM wave states with a certain number of photons $m$.

Then, we can find the matrix elements of the interaction operator of Eq. (24), utilizing the expressions for the DSs

$$\langle \xi_1, m, 2n|\hat{V}_{SR}|\xi_2, m', 2n+1\rangle = \langle \xi_2, m, 2n+1|\hat{V}_{SR}|\xi_1, m', 2n\rangle^*$$
$$= \sum_{\mathbf{k}} \hbar g_{\mathbf{k}} \hat{B}'_k(t) 2u_+^{(m)} u_-^{(m)} i\sin\{2\Omega_0\sqrt{2n+1}t\},$$
$$\langle \xi_2, m, 2n+1|\hat{V}_{SR}|\xi_2, m', 2n+1\rangle = -\langle \xi_1, m, 2n|\hat{V}_{SR}|\xi_1, m', 2n\rangle$$
$$= \sum_{\mathbf{k}} \hbar g_{\mathbf{k}} \hat{B}'_k(t) 2u_+^{(m)} u_-^{(m)} \cos\{2\Omega_0\sqrt{2n+1}t\},$$
$$\hat{B}'_k(t) = \delta_{m',m+1}\hat{b}^+_{\mathbf{k}} e^{-i\delta_{\mathbf{k}}t} + \delta_{m',m-1}\hat{b}_{\mathbf{k}} e^{i\delta_{\mathbf{k}}t}. \quad (27)$$

Here, we assume that the Rabi frequency satisfies the condition $\Omega(m+1) \approx \Omega(m)$ [see Eqs. (4)] in the case of an intensive EM wave with a large number of photons.

The procedure of averaging over the vacuum EM states is performed with the use of the density matrix operator for the heat reservoir [28]

$$\hat{\rho}_R = \prod_{\mathbf{k}}\left(1-\exp\left\{-\frac{\hbar\omega_{\mathbf{k}}}{k_B T_0}\right\}\right)\exp\left\{-\frac{\hbar\omega_{\mathbf{k}}\hat{b}^+_{\mathbf{k}}\hat{b}_{\mathbf{k}}}{k_B T_0}\right\}, \quad (28)$$

where $k_B$ is the Boltzmann constant and $T_0$ is the temperature of the reservoir, which we assume to be zero in the final formulas.

Replacing the sum over discrete wave vectors by an integral over continuous $\mathbf{k}$-space and introducing the spontaneous decay rate in the BS basis, $\gamma_0$ [28], we can define the decay rate between the DSs in Eq. (26) for the reduced density matrix operator as

$$\gamma = \left(u_+^{(m)} u_-^{(m)}\right)^2 \frac{\gamma_0}{2} = \frac{|g_0|^2 (m+1)}{\delta^2 + 4|g_0|^2 (m+1)} \frac{\gamma_0}{2}. \quad (29)$$

It can be seen from Eq. (29) that the decay rate between the DSs depends on the coupling strength of the electron-photon interaction and the detuning of the EM wave frequency from the optical transition between the BSs. Eq. (29) corresponds to the expression for the decay rate in the DS basis obtained earlier [32] without the electron-vibrational interaction.

Then, we obtain the evolution equations for the density matrix elements in the following form

$$\frac{\partial \hat{\rho}_{11}(t)}{\partial t} = -\gamma \hat{\rho}_{11}(t) + \gamma \hat{\rho}_{22}(t),$$
$$\frac{\partial \hat{\rho}_{22}(t)}{\partial t} = -\gamma \hat{\rho}_{22}(t) + \gamma \hat{\rho}_{11}(t),$$
$$\hat{\rho}(t) = \sum_m \hat{\rho}_S^{(m)}(t). \quad (30)$$

This system can be readily solve, and the simple temporal dependence can be obtained for the difference between the diagonal density matrix elements

$$\hat{\rho}_{11}(t) - \hat{\rho}_{22}(t) = (\hat{\rho}_{11} - \hat{\rho}_{22})e^{-2\gamma t}. \quad (31)$$

This result shows that the difference in the electron density, which is initially present at the DSs, decays with the rate $2\gamma$ to the stationary state at which the difference becomes zero (Fig. 3).

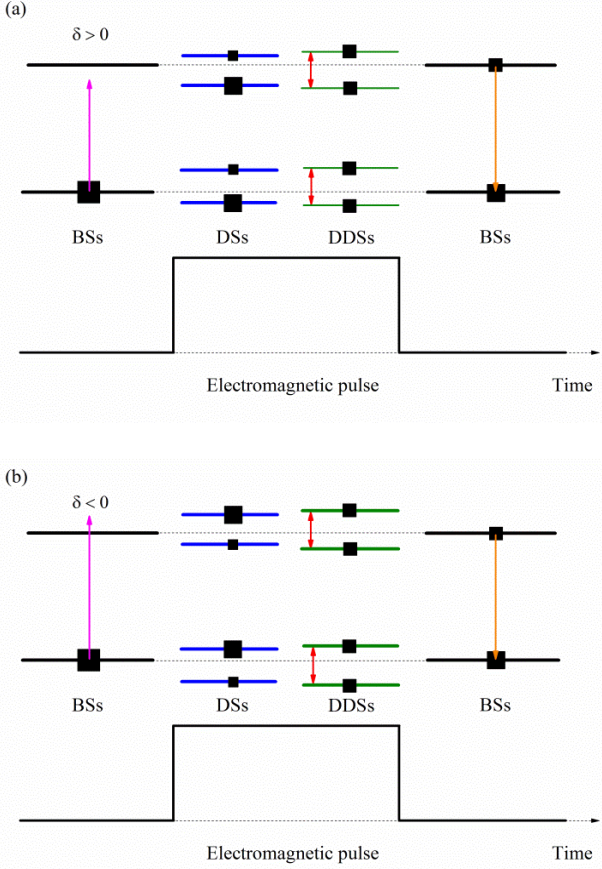

FIG. 3. (Color online) Energy transfer with the use of the dynamical Stark shift in the atomic-like TLS: (a) decreasing of the kinetic energy ($\delta > 0$) and (b) increasing of the kinetic energy ($\delta < 0$).

### B. Mechanism of energy transfer in electron-photon-center-of-mass system

The considered physical processes underlie the mechanism of energy transfer in the electron-photon-center-of-mass system. In the case of fast ROs, a strong EM wave not only rearranges the electron spectrum of an atomic system but also gives rise to the appearance of the VM of the c.m. The existence of the VM in the system provides oscillations of the electron density between the DSs and the establishment of the VC that plays a key role in the energy transfer mechanism. The electron density relaxation between the BSs and DSs due to the spontaneous emission is responsible for the absorption of the EM and vibrational energies, respectively. Finally, the electron populations of the DSs become equal. Alignment of the electron density between the DSs by the use of the interaction of electrons with vibrations of the atomic c.m.

leads to kinetic energy changing of the system. Thus, if initially the lower DS has the larger population than the upper DS, the kinetic energy will be decreased ($\delta > 0$), in the opposite case, if initially the lower DS has the lower population than the upper DS, the kinetic energy will be increased ($\delta < 0$). These cases depict in Fig. 3. The kinetic energy will not be changed if the initial populations of DSs are equal ($\delta = 0$).

The change in the kinetic energy depends on the intensity (Rabi frequency) and frequency (detuning) of the EM wave and is defined as the product of the electron population difference and the energy difference of the DSs in accordance with the energy conservation law

$$2\Omega\left(u_+^2 - u_-^2\right) = \delta . \qquad (32)$$

Thus, we can implement the fast control of atomic kinetic energy by properly choosing the parameters of the laser pulses interacting with a moving atom. In this case, the pulse duration has to be in order of $\gamma^{-1}$ s to complete the population distribution between the DSs. For effective energy transfer in the system, we must use the optical detuning of the same order of magnitude as the resonance Rabi frequency. In this case, the pulse duration is of the order of $0.1/\gamma_0$ s. In addition, the nonadiabatic switching on the interaction of an atom with an EM wave, necessary to trigger ROs, is realized if the switching time (the rise time of the laser pulse amplitude to the maximum value), $\tau$, is much shorter than the reverse of the detuning ($\tau \delta \ll 1$) [29]. Implementing this condition, we depict laser pulses as rectangular ones in Fig. 3. And the considered transference of energy can be observed if the kinetic energy of a free atom will be comparable with the energy of Stark splitting of electron levels.

### IV. CONCLUSIONS

In conclusion, we describe the appearance of the VC in the moving atomic-like system interacting with a strong EM wave. In considered case of the Autler-Townes effect, fast ROs of the electron density give rise to the emerging of the atomic nuclear VM that mixes the DSs and leads to the establishment of the vibronic states in the electron-photon-vibrational system. Then, we calculate the decay rate for the DS basis in the presence of the electron-vibrational interaction and propose the approach of fast kinetic energy control by the use of the laser pulses interacting with the atomic-like system. The result of the interaction depends on the pulse parameters. Thus, the system kinetic energy can be decreased or increased if the detuning between the laser pulse frequency and the optical transition is positive or negative, respectively. The duration of the pulse has to be compared with the inverse of the decay rate for the DS basis, which is the function of the frequency (detuning) and intensity (Rabi frequency) of the laser pulse. For actual values of the detuning, the pulse duration is an order of magnitude longer than the lifetime of the excited atomic level. Additionally, the rectangular pulses must be used for nonadiabatically switching on the electron-photon interaction to trigger the RO regime. Despite its simplicity, the TLS helps to reveal the substantial features of the considered physical mechanism of the VC formation. Nevertheless, the proposed in the paper mechanism can be experimentally implemented for real atoms since it is based on the two well-known for real atomic system effects, namely, the Autler-Townes effect and the optical nutation.

# ACKNOWLEDGMENTS

This work was financially supported by Government of Russian Federation (Grant 08-08) and the Russian Foundation for Basic Research (Grant 17-02-00598).